\documentclass[aps,prl, superscriptaddress, twocolumn, 10pt, floatfix]{revtex4-2}
\usepackage{float}
\usepackage{graphicx}
\usepackage{hyperref}
\hypersetup{colorlinks=true, linkcolor=magenta,citecolor=magenta,urlcolor=magenta}
\begin{document}

% Use the \preprint command to place your local institutional report
% number in the upper righthand corner of the title page in preprint mode.
% Multiple \preprint commands are allowed.
% Use the 'preprintnumbers' class option to override journal defaults
% to display numbers if necessary
%\preprint{}

%Title of paper
\title{Microquasars as the major contributors to Galactic cosmic rays around the \lq\lq knee\rq\rq}

% repeat the \author .. \affiliation  etc. as needed
% \email, \thanks, \homepage, \altaffiliation all apply to the current
% author. Explanatory text should go in the []'s, actual e-mail
% address or url should go in the {}'s for \email and \homepage.
% Please use the appropriate macro foreach each type of information

% \affiliation command applies to all authors since the last
% \affiliation command. The \affiliation command should follow the
% other information
% \affiliation can be followed by \email, \homepage, \thanks as well.
%\author{}
%\email[]{Your e-mail address}
%\homepage[]{Your web page}
%\thanks{}
%\altaffiliation{}
%\affiliation{}
\author{Samy Kaci}
\email{samykaci@sjtu.edu.cn}
\affiliation{Tsung-Dao Lee Institute, Shanghai Jiao Tong University, Shanghai 201210, P. R. China}
\affiliation{School of Physics and Astronomy, Shanghai Jiao Tong University, Shanghai 200240, P. R. China}

\author{Gwenael Giacinti}
\email{gwenael.giacinti@sjtu.edu.cn}
\affiliation{Tsung-Dao Lee Institute, Shanghai Jiao Tong University, Shanghai 201210, P. R. China}
\affiliation{School of Physics and Astronomy, Shanghai Jiao Tong University, Shanghai 200240, P. R. China}
\affiliation{Key Laboratory for Particle Physics, Astrophysics and Cosmology (Ministry of Education) \& Shanghai Key Laboratory for Particle Physics and Cosmology, 800 Dongchuan Road, Shanghai, 200240, P. R. China}

\author{Felix Aharonian}
\affiliation{Yerevan State University, Alek Manukyan St 1, Yerevan 0025, Armenia}
\affiliation{TIANFU Cosmic Ray Research Center, Chengdu, Sichuan, China}
\affiliation{University of Science and Technology of China, 230026 Hefei, Anhui, China}
\affiliation{Max-Planck-Institut für Kernphysik, Saupfercheckweg 1, D-69117 Heidelberg, Germany}

\author{Jie-Shuang Wang}

\affiliation{Max Planck Institute for Plasma Physics, Boltzmannstraße 2, D-85748 Garching, Germany}

%Collaboration name if desired (requires use of superscriptaddress
%option in \documentclass). \noaffiliation is required (may also be
%used with the \author command).
%\collaboration can be followed by \email, \homepage, \thanks as well.
%\collaboration{}
%\noaffiliation

\date{\today}

\begin{abstract}
Recently, LHAASO detected a gamma-ray emission extending beyond $100\,\rm{TeV}$ from 4 sources associated to powerful microquasars. We propose that such sources are the main Galactic PeVatrons and investigate their contribution to the proton and gamma-ray fluxes by modeling their entire population. We find that the presence of only $\sim10$ active powerful microquasars in the Galaxy at any given time is sufficient to account for the proton flux around the knee and to provide a very good explanation of cosmic-ray and gamma-ray data in a self-consistent picture. The $10\,\rm{TeV}$ bump and the $300\,\rm{TeV}$ hardening in the cosmic-ray spectrum naturally appear, and the diffuse background measured by LHAASO above a few tens of $\rm{TeV}$ is accounted for. This supports the paradigm in which cosmic rays around the knee are predominantly accelerated in a very limited number of powerful microquasars.
\end{abstract}

% insert suggested keywords - APS authors don't need to do this
%\keywords{Cosmic-ray Physics --- Gamma-ray Astronomy --- Microquasars}

%\maketitle must follow title, authors, abstract, and keywords
\maketitle

% body of paper here - Use proper section commands
% References should be done using the \cite, \ref, and \label commands
%\section{}
% Put \label in argument of \section for cross-referencing
%\section{\label{}}
%\subsection{}
%\subsubsection{}
%\section{Introduction} \label{sec:intro}
{\it Introduction.} --- Supernova remnants (SNRs) are commonly accepted as effective accelerators of hadronic cosmic rays and have been for a long time argued to be the main Galactic sources of cosmic rays, up to $\rm{PeV}$ energies. However, theoretical studies, e.g. \cite{bell}, have shown that it is very difficult for most SNRs to achieve energies exceeding a few hundreds of $\rm{TeV}$. To circumvent this challenge, it has been suggested that only a small subset of all SNRs can accelerate cosmic rays beyond $\rm{PeV}$ energies with a very hard spectrum, while regular SNRs dominate at lower energies \citep{me}. Despite relaxing the tensions with diffusive shock acceleration theory, this picture still lacks solid observational support. Alternatively, young massive star clusters have been proposed to account for the high-energy end of the cosmic-ray spectrum \cite{ymsc_acc}. However, while particle acceleration to multi-$\rm{PeV}$ energies is possible in some systems, such as Cygnus cocoon \cite{cygnus} and Westerlund 1 \cite{westerlund1}, it still remains strongly dependent on the specifics of each system \cite{clusters} which makes it impossible to reach a definitive conclusion on their viability as the main Galactic PeVatrons. On the other hand, microquasars/X-ray binaries which have been proposed as powerful gamma-ray emitters as early as in the 1990s \cite{felix_1, felix_2, felix_3} have been overlooked for decades and it was not until the recent HAWC, HESS and LHAASO detections of such objects in the ultra-high-energy (UHE) band \cite{lhaasomicro} that the interest of the cosmic-ray community started to shift towards these sources \cite{binaries}.

In this Letter, we investigate, with a self-consistent treatment of cosmic-ray propagation in the Galaxy, a seemingly extreme scenario in which the bulk of $\rm{PeV}$ cosmic rays are contributed by ten or so very powerful microquasars. We show that forecasting powerful microquasars accreting at a rate similar to or higher than their Eddington limit as the main Galactic PeVatrons \cite{microsources} leads to a self-consistent description of the high-energy end of the Galactic cosmic rays. This scenario leads to a very good fit of the cosmic-ray spectrum measured at Earth, with the $10\,\rm{TeV}$ bump and the hardening at $300\,\rm{TeV}$ arising naturally as the footprints of the transition from the regime where supernova remnants are the dominant sources of Galactic cosmic rays to the regime where microquasars such as SS433, GRS1915+105, V4641 Sgr, Cygnus X-3 or Cygnus X-1 dominate the Galactic cosmic-ray flux. In this scenario the cosmic-ray density throughout the Galaxy becomes very clumpy, and the tension between the LHAASO measurement of the ultra-high-energy diffuse gamma-ray background \citep{diffuse_wcda} and theoretical predictions are relaxed. This, in turn, waives the necessity to have a dominant contribution from hypothetical unresolved leptonic gamma-ray sources potentially conflicting with the IceCube measurement from \cite{icecube}.

{\it Methods.} --- We use the Galactic cosmic-ray propagation code presented in \cite{me} allowing for a stochastic injection of particles by individual sources. We generate a population of sources with a birth rate of $0.1\,\rm{kyr}^{-1}$ and a lifetime of $100\,\rm{kyr}$, which leads to a number of $\sim10$ sources present in the Galaxy at any time, which is compatible with the number of high-mass X-ray binaries in \cite{ulx_time, ulx_number}. Regarding the spatial distribution of sources in the Milky Way, we assume that it is axisymmetric and follows the radial distribution of intermediate and high-mass X-ray binaries presented in \cite{space_dist}, and we take a uniform distribution of altitudes with $|z|\leq0.2\,\rm{kpc}$.

For the injection spectrum of cosmic rays we take a power-law of index 2, as suggested by diffusive shock acceleration theory, and an exponential cutoff:
\begin{equation}
    N\left(E,t\right) = \frac{\eta L_0}{\ln{\left(\frac{E_{\rm{max}}}{E_{0}}\right)}}E^{-2}\exp{\left(-\frac{E}{E_{\rm{max}}}\right)}
\end{equation}
where the minimum energy is $E_0 = 1\,\rm{GeV}$, the maximum energy is $E_{\rm{max}} = 10\,\rm{PeV}$, the available luminosity in the system is $L_0$ assumed to follow a power-law of index $-1$ from $10^{39}\,\rm{erg}\,\rm{s}^{-1}$ to $10^{40}\,\rm{erg}\,\rm{s}^{-1}$ assumed to be quasi-constant during the entire lifetime of the source and $\eta = 10\%$ is the fraction of the luminosity that goes to particle acceleration. This choice goes along with the recent LHAASO observations of gamma-ray emitting microquasars, and is notably supported by systems like GRS 1915+105 or SS433 whose jet luminosity can exceed $10^{39}\,\rm{erg}\,\rm{s}^{-1}$ \citep{grs_jets, ss433_jets}.

For cosmic-ray propagation, we numerically integrate the Green function from \cite{me} over the lifetime of the source (treated as point-like), assuming a diffusion coefficient $D\left(E\right)\propto E^{1/3}$ normalized to satisfy the boron to carbon ratio and a halo size $H=3\,\rm{kpc}$ \cite{halo}. In order to compute the gamma-ray emission seen from Earth, we perform an integration over the line of sight and evaluate the gamma-ray production rate through proton-proton collisions on the interstellar gas of \cite{lipari} using the analytical formula given in \cite{gamma}. We account for the impact of heavier nuclei by assuming a nuclear enhancement factor $\epsilon\sim2$ \cite{enhancement} and model the attenuation due to absorption by the photons of the cosmic microwave background (CMB) through an exponential attenuation factor which depends only on the propagation length and the photon energy.

Finally, it should be noted that the number of $\sim10$ active sources is justified by their huge energy budget $\gtrsim10^{38}\,\rm{erg}\,\rm{s}^{-1}$, achievable by only a few Galactic sources, which ensures sufficient cosmic-ray supply in the Galaxy even with a small number of sources. Concerning the spectral index of sources, it is constrained to be hard, close to $2$, in order to guarantee a minimal contribution of microquasars in the $\rm{GeV}-\rm{TeV}$ energy range where SNRs are the dominant sources.

\begin{figure}[h!]
    \centering
    \includegraphics[width=\linewidth]{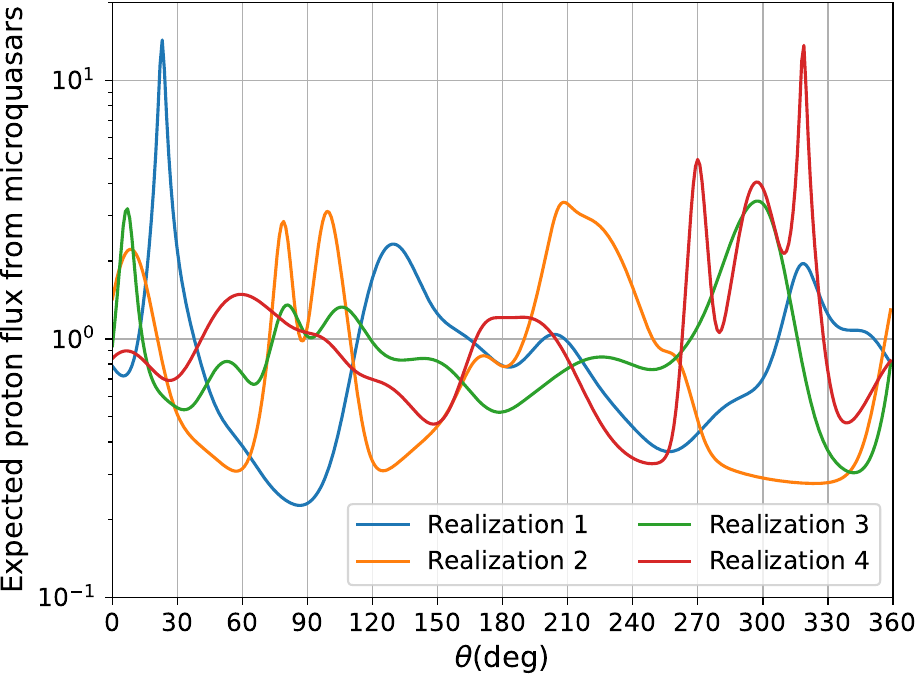}
    \caption{Cosmic-ray proton flux at $3.5\,\rm{PeV}$ (normalized to the Icetop and LHAASO measurements \citep{icetop, lhaaso_proton}) over a circle of radius $8.5\,\rm{kpc}$ centered around the Galactic center for different realizations of the Galaxy. The angle $\theta$ represents different positions on the circle.}
    \label{fig:around}
\end{figure}

{\it The cosmic-ray spectrum.} --- A direct consequence of such a small number of sources is the appearance of strong inhomogeneities in the cosmic-ray density throughout the Galaxy. This is illustrated in figure \ref{fig:around} which displays the proton flux at $3.5\,\rm{PeV}$ (normalized to the Icetop \citep{icetop} and LHAASO \cite{lhaaso_proton} measurements) over a circle of radius $8.5\,\rm{kpc}$ centered at the Galactic center for different realizations of the Galaxy, \textit{i.e.} for different lists of sources drawn randomly.
This figure shows that depending on the current and the recent source activity in the region, the cosmic-ray flux can range from $4$ times smaller than at the Earth location (long time without any source) to more than one order of magnitude higher than the flux at the Earth location (in the vicinity of an active source). In this context, assuming that the Earth is in a region of low cosmic-ray density would provide a natural interpretation for the discrepancy between the expectations from the target gas distribution and the measurement reported by LHAASO in \cite{lhaaso_diffuse}.

In this context, inferring the contribution of powerful microquasars to the cosmic-ray flux at Earth is no longer straightforward and needs to be done cautiously. As a part of this investigation, we first evaluate the contribution that would come only from the diffuse sea of cosmic rays contributed by the bulk of old inactive sources. This can be achieved by generating lists of sources that do not contain any source younger than $500\,\rm{kyr}$ with a distance to Earth smaller than $4\,\rm{kpc}$. This choice of parameters is made to coincide with the residence time of $3.5\,\rm{PeV}$ cosmic rays, around $\sim650\,\rm{kyr}$. At this time the cosmic rays of such sources will have diffused over a distance of $\sim4\,\rm{kpc}$. Although this scenario is not likely to happen with our simulation setup (probability around $1\%)$, it can still serve to extract a lower limit, as it is representative of a region with a very small recent source activity. As a self-consistency check one can \textit{a priori} estimate this lower limit to be around $\sim25\mbox{--}50\%$ of the flux reported by Icetop based on the minima in figure \ref{fig:around}. The second step is to evaluate the expected contribution of microquasars at the Earth location, which is performed by removing the restrictions introduced earlier in the generation of the source lists.

\begin{figure}
    \centering
    \includegraphics[width=\linewidth]{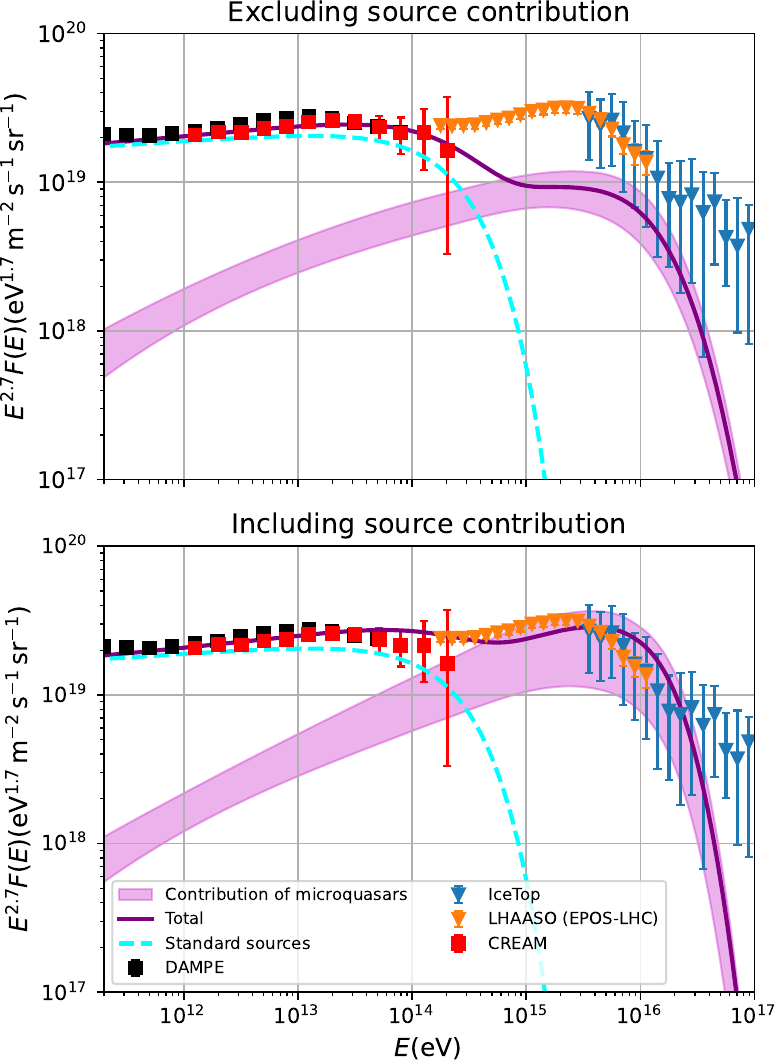}
    \caption{Contribution of microquasars to the proton spectrum. In both panels, the dashed cyan curve shows the flux from a standard population of Galactic sources producing a power-law spectrum with an exponential cutoff at $260\,\rm{TeV}$, the magenta shaded area shows the flux contributed by microquasars assumed to produce a power-law spectrum of index $2$ with an exponential cutoff at $10\,\rm{PeV}$ and the purple curve represents the sum of the cyan and the magenta curves. The upper panel shows the sole contribution of the sea of cosmic rays due to miroquasars and the lower panel shows the total contribution of microquasars. The data points are from DAMPE \citep{dampe}, CREAM \citep{cream}, Icetop \citep{icetop} and LHAASO \cite{lhaaso_proton}.}
    \label{fig:fit}
\end{figure}

In figure \ref{fig:fit}, the upper panel shows the contribution solely originating from the sea of cosmic rays from microquasars, while the lower panel shows the total contribution of microquasars, including the potential contribution of current and/or recent sources. The contribution of microquasars is shown as the magenta shaded area which in both panel represents the $68\%$ containment region. We report the results over $100$ ($10^4$) simulations for the upper (lower) panel. 

From the upper panel, it is clear that, above the knee, microquasars must have an important contribution to the cosmic-ray flux, which is at least at the level of $25\%$ of the flux measured by Icetop and LHAASO, while due to the very hard injection spectrum of cosmic rays, this contribution is much less important but still not negligible at lower energies. Moreover, looking at the lower panel, we observe that microquasars have the potential to completely contribute to the cosmic-ray flux around the knee and induce the hardening observed by LHAASO \citep{lhaaso_proton} and GRAPES-3 \citep{grapes} around $300\,\rm{TeV}$, while an early cutoff of the standard population of sources, assumed to be SNRs (see dashed cyan curve in figure \ref{fig:fit}), would provide a very good fit to the $10\,\rm{TeV}$ bump \cite{tev_bump}. This result is important because it does not involve the presence of a hypothetical local source \cite{binaries} or local shocks reaccelerating cosmic rays at a few parcecs from the Earth \cite{shock_reac}. A scan of the parameter space reveals that in order to fit the data, an efficiency of $\sim10$ to $\sim20\%$ and an injection index around $1.9$ to $2.05$ are required for particle acceleration in microquasars, which justifies \textit{a posteriori} our choice, while the cutoff energy of the standard source population must be between $150\,\rm{TeV}$ and $300\,\rm{TeV}$ close to the maximum energy in most supernovae \citep{bell}. Softer microquasar spectra require higher efficiencies and lower SNR energy cutoffs. Finally, it is worth mentioning that for the lower panel some simulations led to a cosmic-ray flux more than $2$ orders of magnitude higher than the flux measured at Earth, which is consistent with our interpretation of Figure \ref{fig:around}.

{\it Gamma-ray counterpart.} --- While cosmic rays only allow to extract local information, gamma-ray observations allow to see most of the Galaxy for energies smaller than a few hundreds of $\rm{TeV}$ and can be used to investigate the properties of the sources and the propagation of their cosmic rays in the interstellar medium. Here, we quantify the contribution of microquasars to the diffuse gamma-ray background measured by LHAASO.

\begin{figure}
    \centering
    \includegraphics[width=\linewidth]{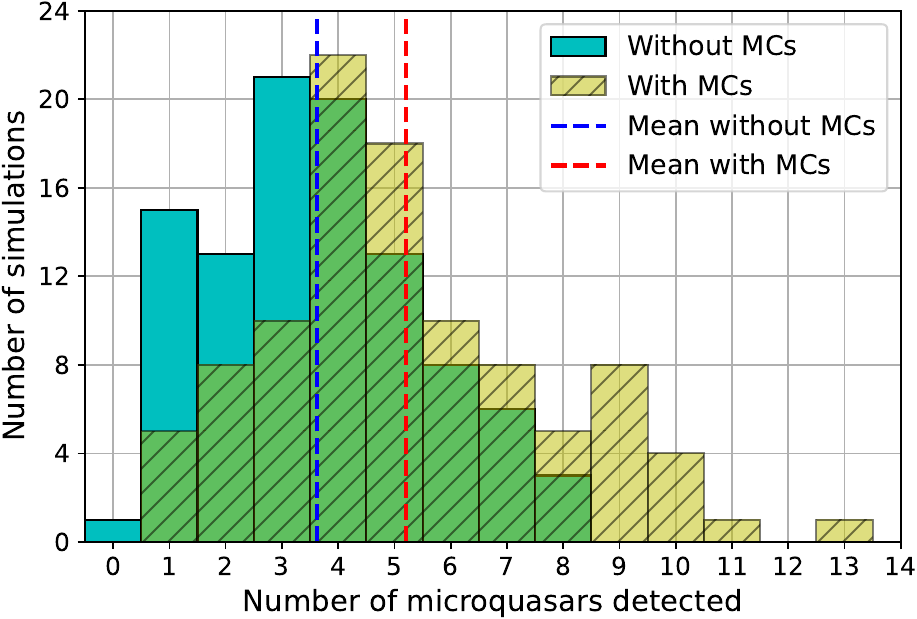}
    \caption{Histogram displaying the number of microquasars detected in the FOV of LHAASO  at $100\,\rm{TeV}$ per simulation over $100$ simulations. The blue histogram represents the number of detections without MCs and the yellow dashed histogram represents the number of detections with molecular clouds with a density $n=30\,\rm{cm}^{-3}$ and a radius $r=50\,\rm{pc}$. The blue (red) vertical dashed line represents the average number of detections without (with) MCs.}
    \label{fig:number_detected}
\end{figure}

In order to estimate the contribution of microquasars to the diffuse background of LHAASO we first remove the contribution of those sufficiently bright to be detected by LHAASO. In order to investigate the detectability of sources, we mainly follow the approach of \cite{amato}. To compute the gamma-ray flux, we use the target gas density and cosmic-ray distribution presented in \cite{lipari}. In figure \ref{fig:number_detected} we present our estimate, over 100 simulations, of the number of microquasars that would be detected in the FOV of LHAASO with and without the help of molecular clouds (MCs) that may be present in the vicinity of the sources and, in both cases, require a significance of $5\sigma$ to claim a detection. For the former case, we assume that all the MCs can be modeled as uniform spheres assuming different combinations of densities and radii ranging respectively between $30\,\rm{cm}^{-3}$ and $300\,\rm{cm}^{-3}$ and between $20\,\rm{pc}$ and $50\,\rm{pc}$. Note that the latter condition corresponds to a total mass of $\sim10^{6.5}\,M_{\odot}$ close to the maximum mass of molecular clouds in our Galaxy \cite{mcs}.

For all cases, we observe that the number of detectable sources peaks around a few with the mean number of detectable sources located around $3.6$ without the help of MCs whereas it ranges between $5.1$ and $5.6$ for all MC densities and radii. This result, compatible with the observations of LHAASO \citep{lhaasomicro}, further supports our scenario.

\begin{figure}
    \centering
    \includegraphics[width=\linewidth]{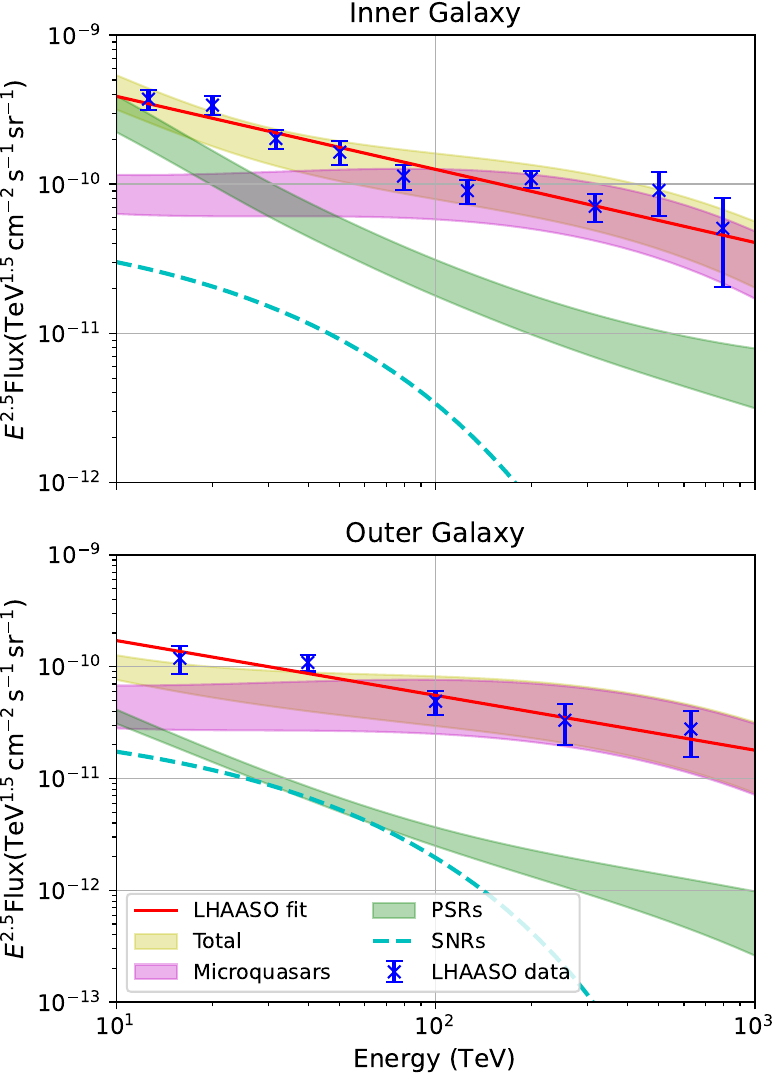}
    \caption{Contribution of microquasars to the Galactic diffuse gamma-ray emission reported by LHAASO \cite{lhaaso_diffuse}. The upper (lower) panel shows the data related to the inner (outer) Galaxy defined as the region $15^{\circ}\leq l<125^{\circ}$ and $\left|b\right|\leq5^{\circ}$ ($125^{\circ}\leq l<235^{\circ}$ and $\left|b\right|\leq5^{\circ}$). The contribution of unresolved pulsars shown in green is taken from \cite{me2}. In both panels, the yellow area represents the sum of the contributions of microquasars (magenta area), pulsars (green area) and SNRs (dashed cyan curve) assumed to have a cutoff at $200\,\rm{TeV}$. The shaded areas represent one standard deviation over $100$ simulations.}
    \label{fig:diffuse}
\end{figure}

Once all detected sources have been removed, the contribution of microquasars to the UHE diffuse gamma-ray background measured by LHAASO can be determined by adopting the same masking procedure as LHAASO in \cite{lhaaso_diffuse}. This contribution will henceforth include that of the sea of hadronic cosmic rays accelerated by microquasars and that of unresolved microquasars. Figure \ref{fig:diffuse} displays the measurement of the diffuse gamma-ray background by LHAASO \cite{lhaaso_diffuse} for the inner and outer Galaxy as defined in \cite{diffuse_wcda, lhaaso_diffuse}, together with our expected diffuse emission originating from microquasars (magenta shaded area), SNRs (cyan dashed curve), and unresolved pulsars (green shaded area) presented in Ref. \cite{me2}.

Figure \ref{fig:diffuse} shows that the sole contribution of microquasars together with that of the unresolved pulsar population as derived in \cite{me2} is sufficient to explain the UHE data of LHAASO. Above $50\,\rm{TeV}$, microquasars can account for almost all the gamma-ray flux with a minimal contribution from unresolved pulsars in the inner Galaxy, whereas in the outer Galaxy, microquasars alone are sufficient to explain the measurements of LHAASO above $10\,\rm{TeV}$. This result is in agreement with our previous findings \cite{me2} and is of great significance. It notably waives the need to have an important yet speculative contribution from unresolved pulsars \cite{neutrinos, hooper}, which are the main leptonic PeVatrons, and prevents tensions with neutrino observations \cite{icecube}.

{\it Discussion.} --- In this Letter, we have investigated the observational features arising from the forecasting of a small number of powerful microquasars as the main Galactic PeVatrons for hadronic cosmic rays.

As part of our assumptions, we have considered a lifetime of $100\,\rm{kyr}$ for all microquasars with a birth rate of $0.1\,\rm{kyr}^{-1}$ based on \cite{ulx_time} and do not expect our results to be sensitive to variations in the lifetime of sources and their birthrate within the ranges we are interested in. The number of $\sim10$ sources simultaneously present in the Galaxy is supported by the observations of LHAASO \cite{lhaasomicro} and is already very close to the minimum number of sources required in the Galaxy \cite{max_energy}. As a result, one could reduce the birthrate of sources only if their lifetime is increased accordingly, which would further enhance the spatial fluctuations in the cosmic-ray density. Alternatively one could increase the birthrate of sources while decreasing their lifetime, which would yield results similar to those in \cite{me}. A birthrate of $\gtrsim0.7\,\rm{kyr}^{-1}$ would, for instance, correspond to the case of $\sim2\%$ of PeVatron SNRs in \cite{me}. 

Another scenario not presented in this Letter is the case of softer injection spectra, such as a spectral index of $2.2$ usually expected for particle acceleration at relativistic shocks \cite{relat_shock1, relat_shock2}. A good fit of the data is also possible in this case and only requires a smaller contribution from SNRs and a higher contribution from microquasars. The latter can be obtained by increasing the efficiency of particle acceleration to $\sim40\%$, a higher source birthrate or simply a less frequent, but not fully unrealistic, configuration of the Galaxy where more sources are present close to Earth, leading to an increased cosmic-ray flux.

Finally, we ignored the case where different sources would have different $E_{\rm{max}}$. In that case, the cosmic-ray spectrum would be dominated by the closest source if it is sufficiently close, alternatively, it would appear as the average spectrum of the cosmic-ray sea.

{\it Conclusion.} --- In this Letter, we have demonstrated that assuming a small number of $\sim10$ very powerful PeVatron microquasars, where the compact object accretes at a rate close to or above the Eddington limit, can provide a natural interpretation to both cosmic-ray and gamma-ray data in a self-consistent picture. In this picture, SNRs would be the main low-energy cosmic-ray sources, accelerating particles up to the energy of $\sim200\,\rm{TeV}$, above which microquasars, which are capable of accelerating particles with harder spectra, would take over. The $10\,\rm{TeV}$ bump in the cosmic-ray spectrum and the $300\,\rm{TeV}$ hardening would arise as the footprints of this transition, while the diffuse background measurements of LHAASO above a few tens of $\rm{TeV}$ would be well fitted without invoking a dominant contribution from a speculative population of unresolved pulsars. Near-future gamma-ray observations may reveal more UHE emitting microquasars further supporting our findings and providing the confirmation that microquasars are the main Galactic hadronic PeVatrons.

{\it Acknowledgments.} --- Samy Kaci acknowledges funding from the Chinese Scholarship Council (CSC) and thanks Ramiro Torres-Escobedo for his valuable help and support. Felix Aharonian acknowledges support from the Sichuan Science and Technology Department (under grant number 2024JDHJ0001). This work is supported by the National Natural Science Foundation of China under Grants Nos. 12350610239, and 12393853. This work was supported by the National Center for High-Level Talent Training in Mathematics, Physics, Chemistry, and Biology.

\end{document}